\documentclass{icrc2009}

\usepackage{graphicx}   % for including figures
\usepackage[font=footnotesize,caption=false]{subfig} % subfig.sty for a double column floating figure using two subfigures
\usepackage{fixltx2e}
\usepackage{url}

\newcommand{\shorttitle}[1]%
{\markboth{Proceedings of the 31\MakeLowercase{$^{st}$} ICRC, {\L}\'{o}d\'{z} 2009}{#1} }
\newcommand{\etal}{\MakeLowercase{\textit{et al. }}} % "et al."

\begin{document}
\title{Physics Capabilities of the IceCube DeepCore Detector}

\author{\IEEEauthorblockN{Christopher Wiebusch\IEEEauthorrefmark{1}
			 for the IceCube Collaboration\IEEEauthorrefmark{2}
                           } \\
\IEEEauthorblockA{\IEEEauthorrefmark{1}III.Physikalisches Institut, RWTH Aachen, University, Germany}
\IEEEauthorblockA{\IEEEauthorrefmark{2} See the special section of these proceedings}
}

% please write the preseter's name and short title (3-4 words maximum)
%    which will appear at the header of the even pages.
\shorttitle{Christopher Wiebusch \etal  Capabilities of  IceCube DeepCore}
\maketitle

\begin{abstract}
 IceCube-DeepCore  is a compact Cherenkov detector located in the 
clear ice of the bottom center of the IceCube Neutrino Telescope. 
Its purpose is to enhance the sensitivity of IceCube for low neutrino 
energies ($\mathbf{<1}$\,TeV) and to lower the detection threshold of IceCube by 
about an order of magnitude to below 10\,GeV. The detector is formed by 
6 additional strings of 360 high quantum efficiency phototubes 
together with the 7 central IceCube strings. The improved 
sensitivity will provide an enhanced sensitivity to 
%WIMP annihilations
% in the Sun and in the Earth and 
probe a range of parameters of dark  matter models not covered by direct 
experiments. It opens a 
new window for atmospheric neutrino oscillation measurements 
of $\mathbf \nu_\mu $ disappearance or $\mathbf \nu_\tau $ appearance in an energy region 
not well tested by previous experiments, and enlarges the field of view 
of IceCube to a full sky observation when searching for potential neutrino 
sources. The first string was succesfully installed in January 2009, commissioning
 of the full detector is planned early 2010.
% after the installation of 
%the remaining $5$ strings. 
  \end{abstract}

 
\section{Introduction}

Main aim of the IceCube neutrino observatory \cite{IC3}
 is the detection of high 
energy extraterrestrial neutrinos from cosmic sources,
 e.g. from active galactic nuclei. 
%The detection of high energy neutrinos would help to identify the 
%sources and the acceleration
%mechanisms of high energy cosmic rays, which has been an 
%unresolved question for
%almost a century. 
The detection of high energy neutrinos would help to resolve the
question of  the 
sources and the acceleration
mechanisms of high energy cosmic rays.

IceCube is located at the geographic South Pole.
The main instrument of IceCube %, \emph{InIce}, 
will consist of $80$ 
cable strings, each with $60$
highly sensitive photo-detectors which are installed in the 
clear ice at depths between
$ 1450 $\,m and $2450$\,m below the surface. 
Charged leptons with an energy above $100$\,GeV inside or close to the 
detector produce enough Cherenkov light to be detected and
reconstructed using the timing information of the photoelectrons 
recorded with large area phototubes (PMT).
While the primary goal is of highest scientific interest, 
the instrument can address a multitude of scientific questions, 
ranging from fundamental physics 
 such as %dark matter and 
physics on energy-scales beyond
the reach of current particle accelerators to multidisciplinary
aspects e.g.\ the optical properties of the deep Antarctic ice which reflect climate changes
on Earth.

IceCube  is complemented by other major detector components. 
The surface air-shower detector \emph{IceTop} is used to  study  
high energy cosmic rays and to  calibrate \emph{IceCube}. 
R\&D studies are underway to supplement IceCube with radio (\emph{AURA}) and acoustic sensors (\emph{SPATS}) in order to extend the energy range
beyond EeV energies.
DeepCore will consist of six additional densely instrumented strings
 deployed in the bottom center of
 IceCube  and is the focus of this paper.
%Six additional and more densely instrumented strings 
%will be deployed in the bottom center of
%the IceCube detector and form the here considered \emph{DeepCore} detector.

The first DeepCore string has been taking data since its successful installation in January 2009.
%A first DeepCore string has been succesfully 
%installed in January 2009 and is taking data
%since then. 
The DeepCore detector will be completed in 2010
and will replace the existing \emph{AMANDA-II} detector, which was 
decomissioned in May 2009.
DeepCore will lower the 
detection threshold of IceCube by an order of magnitude to below 
$ 10 $\,GeV and, due to its improved design, 
provide new capabilities compared to AMANDA.
In this paper we describe the design of  DeepCore and the 
 enhanced physics capabilities that may  be addressed.

\section{DeepCore Design and Geometry}

 % see \section{Examples of \LaTeX\  instructions} and \subsection{Figures}
  \begin{figure}[!t]
   \centering
   \includegraphics[width=2.5in]{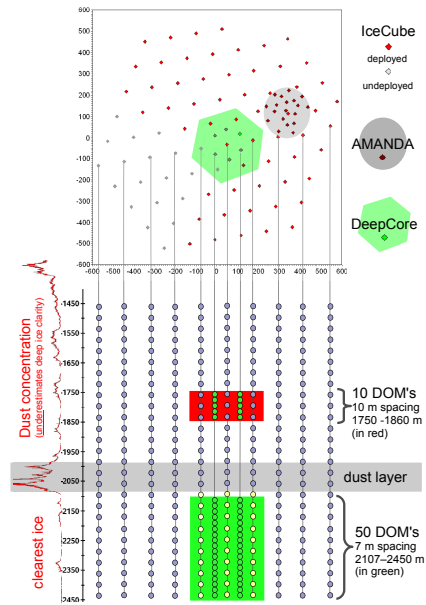}
   \caption{The geometry of the  DeepCore Detector. 
The top part shows the surface projection of horizontal
 string positions 
and indicates the positions of AMANDA and DeepCore. 
The bottom part indicates the depth of sensor  positions.
At the left  the depth-profile of the optical  transparency 
of the ice is shown.
 }
   \label{fig:dc_geo}
  \end{figure}

The geometry of DeepCore is sketched in figure \ref{fig:dc_geo}.

DeepCore is comprised  of $6$  additional 
strings, each of which are instrumented with $60$ phototubes, in conjunction
with the $7$ central IceCube strings.
The detector is divided into two components.
Ten sensors of each new string are at  shallow depths between
$ 1750 $\,m and $1850$\,m,  above a major dust-layer
of poorer optical transparency and will be used as a 
veto-detector for the deeper component.
The deep component is formed by $50$ sensors on each string and is
 installed in the  clear ice at depths between
$ 2100 $\,m and $2450$\,m. % below the surface.
It will  form, together with the neighboring  IceCube sensors
%within that particular volume, 
the main physics volume.

The deep ice  is on average % 40\%-50\% clearer than
twice as clear as 
the average ice above $2000$\,m \cite{IC3ice}. 
The effective scattering length 
reaches $50$\,m and the absorption length $230$\,m.
Compared to AMANDA a substantially 
larger number of unscattered photons will be recorded allowing for an
improved pattern recognition and reconstruction of
 neutrino events in particular at lower energies.
%for  the smaller light yield at lower energies.

Another important aspect % of the design 
is a denser spacing of photo-sensors
compared to IceCube:
The horizontal inter-string spacing is $72$\,m  (IceCube: $125$\,m).
%The  corresponding average
%distance of events to sensors is thus smaller than the  scattering length.
The vertical spacing of sensors along a string  is only $7$\,m 
(IceCube: $17$\,m).

  \begin{figure}[!t]
   \centering
   \includegraphics[width=2.5in]{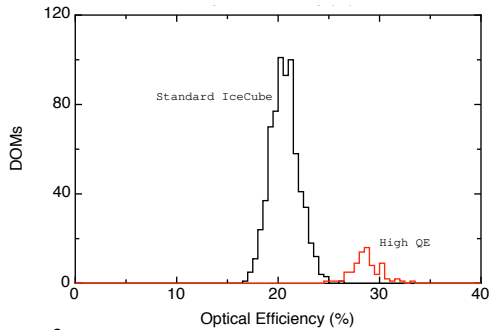}
   \caption{Results of the quantum efficiency calibration at $\lambda=405 $\,nm
of the  DeepCore phototubes compared to standard IceCube phototubes.}
   \label{fig:dc_pmt}
  \end{figure}

The next major improvement with respect to IceCube and AMANDA is the 
usage of new phototubes (HAMAMATSU R7081-MOD) of higher quantum efficiency.
This hemispherical 10'' 
PMT is identical to the standard IceCube PMT
\cite{IC3dom}, but employs a modified cathode material of higher
%which improves the 
quantum efficiency (typically  $33$\,\% at $\lambda = 390$\,nm).
%from typically  $25$\,\% to $33$\,\% at $\lambda = 390$\,nm.
%A disadvantage is an enhanced noise rate of 33\% at $-45^\circ$\,C, compared
%to standard IceCube sensors.
Calibrations of the phototubes for DeepCore confirm a sensitivity
 improvement
of $30$\,\%-$40$\,\% with respect to the standard IceCube PMT 
(figure \ref{fig:dc_pmt}).
%It is forseen, that 
Also regular IceCube strings will be equipped 
with these phototubes within the DeepCore  volume.

The net effect of the denser instrumentation %for the considered volume 
is a  factor $\sim 6$ gain in sensitivity
%with respect to 
for photon detection and  superior  optical clarity of the ice.
This is an imporant  prerequisite for a substantially lower
detection threshold.
% and the anticipated physics goals.

\section{DeepCore Performance}

The electronic hardware %and construction 
of the optical sensors is identical
to the standard IceCube module \cite{IC3dom} and this significantly 
reduces the efforts for
 maintenance and operations compared to AMANDA.
The DeepCore detector is integrated 
into a homogeneous data aquisition model of IceCube which will be 
only supplemented by an additional trigger. % and data selection algorithms.
Initial comissioning data  of the first installed DeepCore string
verifies that the hardware works reliably and as expected.

The IceCube detector  is triggered if typically 
a multiplicity of $8$ sensors within $\sim 5\,\mu$s  observe  
a signal coincident with a hit in a 
neighboring or next to neighboring sensor.
For each 
trigger, the   signals of the full detector are  
transferred to the surface.

\begin{figure}[!t]
   \centering
   \includegraphics[width=2.8in]{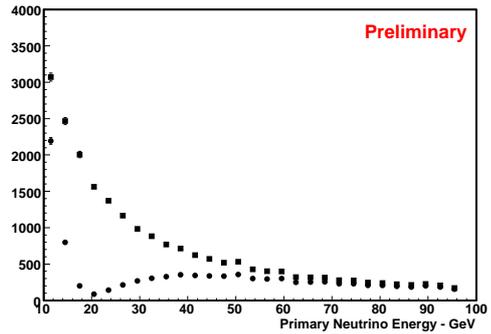}
   \caption{Number of triggered vertical atmospheric neutrinos per year (per 3GeV) versus the neutrino energy. 
Events from $1.6 \pi $\,sr are accepted.
 Shown are the numbers without (squares) and with (circles) 
%(SMT-4),
the inclusion of oscillations ($\Delta m_{atm}^2 = 0.0024$\,eV$^2$, $\sin(2\theta_{23})=1$).}
   \label{fig:osci}
\end{figure}

For the  sensors within  the 
considered volume  the data taking is supplemented  with a 
reduced multiplicity requirement of typically $3-4$.
 As shown in figure \ref{fig:osci}, 
such a trigger is sufficient to 
trigger atmospheric neutrino events down to a 
threshold of $1$\,GeV, sufficiently below the anticipated physics threshold.

The chosen location of DeepCore allows to utilize the 
outer IceCube detector as an active veto shield against the background of
down-going atmospheric muons. These  are detected at a $ \sim 10^6$ 
higher rate than neutrino induced muons.
The veto provides  external information  to suppress this background
and standard up-going neutrino searches will strongly benefit from a 
larger signal efficiency and a lower detection threshold as the demands
on the maturity of recorded signals decrease.

Even more intriguing is the opportunity to identify down-going 
$\nu $ induced $\mu $, which may, unlike cosmic ray 
induced atmosperic $\mu $,  start inside the DeepCore detector.
Simulations \cite{IC3veto} show that three rings of surrounding IceCube strings
and the instrumentation in the upper part of IceCube are sufficient
to achieve a rejection of atmospheric muons by a factor $> 10^6$
maintaining a large fraction of the triggered neutrino signals.
A further  interesting aspect is the proposal  \cite{nuveto} to  
veto also  atmospheric $\nu $ by the detection of a correlated atmospheric 
$\mu $. This could provide the opportunity to reject a substantial part of 
this 
usually irreducable background for extra-terrestrial neutrino searches.

Triggered events which start inside the detector will be selected online 
 and transmitted north by satellite. 
 Already simple algorithms 
% which are based on the location and the arrival times 
allow to suppress the background rate by a factor $> 10^3 $ 
and meet the bandwidth requirements 
while 
keeping 90\% of the signal \cite{IC3veto}.
%Main requirement is to reduce the background by at least 2-3 orders 
%of magnitude to meet the bandwidth requirements 
%while keeping a mximum efficiency for 
%potential signal events.
A typical strategy requires that the earliest hits are  
located inside DeepCore
and allows for  later hits in the veto-region only if 
the time is causally consistent with the hypothesis of a starting track.
A filter which selects starting tracks in IceCube has been  active since
2008 and allows performance verification of such filters with experimental 
data and to benchmark the subsequent physics analysis.

The filtered events are analyzed offline 
with more sophisticated reconstruction 
algorithms. Here, the focus is to improve the purity of the sample and to 
reconstruct direction, energy and the position of the interaction vertex.
A particularily efficient likelihood algorithm ({\texttt{finiteReco}} \cite{IC3veto}) 
capable of selecting  starting muons  evaluates the 
hit probabilities of photomultipliers with and without a signal in dependence
of the distance to the track.
It estimates the most probable position of the start-vertex and  
 provides the probablity that a track may have reached this point
undetected by the veto.

 \begin{figure}[!t]
   \centering
%preliminary
   \includegraphics[width=2.5in]{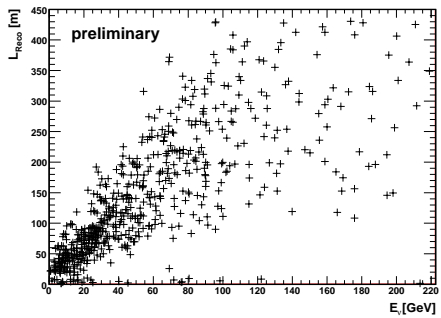}
   \caption{The reconstructed length of $\mu $ contained tracks in 
DeepCore,  based on the reconstructed start- and stop- 
vertex with the {\texttt{finiteReco}} algorithm.
The data are  $\nu$ induced $\mu$-tracks from the upper hemisphere, 
which are reconstructed to start within DeepCore.}
   \label{fig:dc_length}
  \end{figure}

The reconstruction algorithms are still under 
development but initial results are 
promising. As an example, figure \ref{fig:dc_length} 
shows the reconstructed length of $\mu $ tracks as function of the 
$\nu $ energy. Already the currently achieved resolution of $\sim 50$\,m 
results in a visible correlation with the neutrino energy in particular 
for energies $\le 100$\,GeV. Note, that the resolution  is 
substantially better for vertical tracks.

 \begin{figure}[!t]
   \centering
   \includegraphics[width=2.8in]{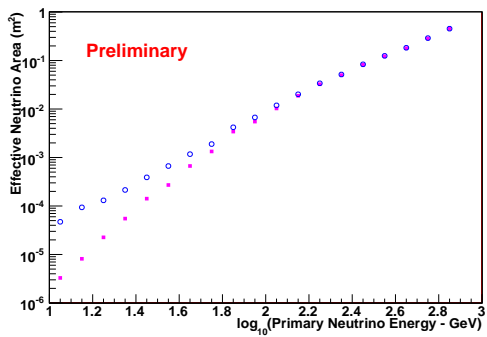}
   \caption{Effective neutrino detection area of IceCube (trigger level) versus the energy for up-going neutrinos. 
The squares are IceCube only. The circles represent  the area 
if  DeepCore is included.}
   \label{fig:dc_area}
  \end{figure}

The effective detection area  of IceCube for neutrinos for  triggered events
 is  shown in figure \ref{fig:dc_area}. 
Despite DeepCore being  much smaller  than IceCube,
 a substantial gain of up to an order of magnitude 
 is achieved by the additional events detected in DeepCore. 
Higher level event selections for specific  physics analysis  
benefit strongly from the higher 
 information content of events and
the gain of DeepCore further  improves.

\section{Physics Potential}

\subsection{Galactic point sources of neutrinos}

\begin{figure}[!t]
   \centering
   \includegraphics[width=3.35in]{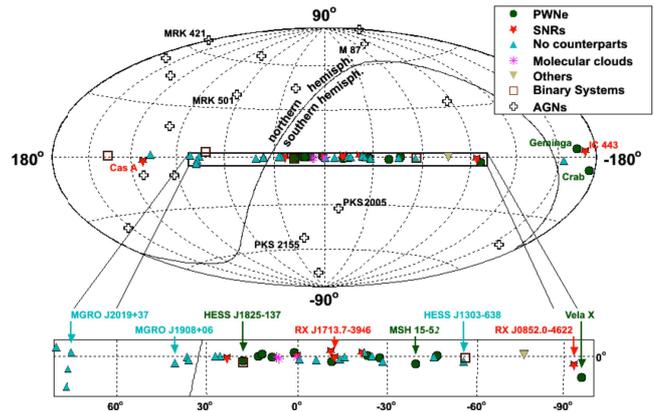}
   \caption{Interesting celestrial objects with known emission of TeV gamma 
rays. 
%Shown are the positions in galactic coordinates with a 
%particular highlight on the survey of the galactic plane, 
%performed by the H.E.S.S. instrument.\cite{HESS}
}
   \label{fig:sources}
  \end{figure}

The analysis of IceCube data greatly benefits from the location 
at the geographic South Pole because  the celestial 
sphere fully rotates during one sideral 
day. Azimuthal detector effects are largly washed
out because each portion of the sky is observed 
with the same exposure and same inclination.
 However, the  aperture of the conventional up-going muon 
analysis is restricted 
to only the Northern hemisphere and  leaves out a large fraction of the 
galactic plane and a  number of interesting  objects 
such as the galactic center (see figure \ref{fig:sources}).
Extending the field of view of IceCube 
at low energies ($\le 1$\,TeV) to a full sky observation 
will greatly enlarge the number of interesting galactic sources 
in reach of IceCube\footnote{Note, that at high energies $> 1$\,PeV 
the background 
 of atmospheric muons rapidly decreases and also here neutrinos 
 from the Southern hemisphere can be detected by IceCube 
 \cite{hepointdown}. However, galactic sources are usually 
 not expected to produce significant fluxes of neutrinos 
 at energies around the cosmic ray knee and above.} 
.

The energy spectrum %observation 
of   gamma rays from supernova remnants show indications 
of a potential cut-off %in the energy spectrum of 
at a few TeV \cite{HESS}. 
Under the assumption of a  hadronic production mechanism 
for these  gamma rays the corresponding neutrino fluxes would 
show a similar cut-off at typically half of that cut-off  value.
The high sensitivity of DeepCore for neutrinos of TeV energies and below 
 will  complement the  sensitivity of IceCube which is optimized for  energies 
of typically $10$\,TeV and above.

\subsection{Indirect detection of dark matter}

The observation of an excess of high energy neutrinos from the direction 
of the Sun can be interpreted by means of annihilations of 
WIMP-dark matter in its center.
 The energy of such neutrinos is  a fraction of the mass of the WIMP
particles (expected on the TeV-scale) and it depends  on the decay chains
 of the annihilation products.
The large effective area of DeepCore and the 
possibility of a highly  efficient signal selection
greatly improves the sensitivity of IceCube. 
In particular it is possible to probe
regions of the parameter space
with soft decay chains and  WIMP masses below $\sim 200 $\, GeV and 
which are not disfavored by direct search experiments.

\begin{figure}[!t]
   \centering
%preliminary
   \includegraphics[width=2.9in]{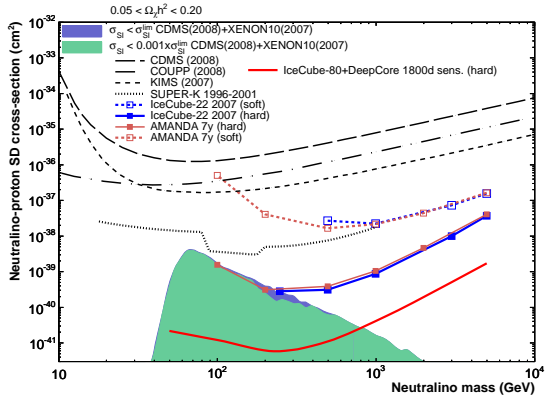}
   \caption{The expected upper limit of IceCube DeepCore at 90\% 
 confidence level on the
     spin-dependent neutralino-proton cross section for the hard
     (W$^+$W$^-$) annihilation channel as a function of the neutralino
     mass for IceCube  including Deep
     Core (solid  line). Also shown are limits from previous direct and indirect searches. The shaded areas represent MSSM models which are not
     disfavoured by direct searches, even if their sensitivity would be improved by a factor 1000.
%The limits from CDMS [15], COUPP [21], KIMS[20] and Super-K [17] are shown for
%     comparison.  
}
   \label{fig:wimp}
\end{figure}

An example of the sensitivity
for the hard annihilation
channel of supersymmetric neutralino dark matter
 is 
\nopagebreak
shown 
in  figure \ref{fig:wimp}.

\subsection{Atmospheric neutrinos}

DeepCore will trigger on the order of 
$10^5 $ atmospheric neutrinos/year in the energy range 
from $1$\,GeV to $100$\,GeV. 
Atmospheric neutrinos are largly unexplored in this energy range.
Smaller experiments like Super-Kamiokande
cannot efficiently measure the spectrum for energies above $10 $\,GeV
and measurements done by  AMANDA  only 
start at $ 1$\,TeV.
In the range between  $30-50$\,GeV decays of charged kaons 
become dominant over decays
of charged pions \cite{KAON} for the production of atmospheric neutrinos
and  the systematic error of flux  calculations 
 increases. A measurement of  this 
transition could help to reduce systematic errors of the 
flux of atmospheric neutrinos at TeV energies.

The first maximum of disappearence of atmospheric $\nu_\mu $ due to 
 oscillations appears at  an energy of about $25 $\,GeV for vertically 
up-going atmospheric neutrinos \cite{DARREN}.
The  energy threshold of about $10$\,GeV   
 would allow to measure atmospheric neutrino oscillations by means of a direct 
observation of the oscillation pattern in this energy range.
In addition, DeepCore would aim to observe the  
appearance of $\nu_\tau $  by the detection
of small cascade-like events in the DeepCore volume at a rate which is 
anti-correlated with the  disappearance of $\nu_\mu$.

Similar to $\nu_\tau$, the signature of $\nu_e $ events are cascade-events
 with a large local light deposition without the signature of a track.
The dominant background to these events are charged current $\nu_\mu $ interactions with a small momentum transfer to the $\mu $. Analyses like 
these will have to be performed considering all three flavors and their mixing.
Note, that only for a further reduction of the energy threshold smaller than $ 10$\,GeV 
matter effects in the Earth's core would become visible \cite{DARREN}.

\subsection{Other physics aspects}

Two remaining items are only briefly mentioned here.

Slowly moving magnetic monopoles, when  catalizing proton decays,  produce
subsequent energy depositions of $\sim 1$\,GeV along their path 
with  time-scales of $\mu$s to ms. 
Initial studies are under-way to develop a dedicated 
 trigger for this  signature using  delayed coincidences.

DeepCore extends the possibility to search for neutrino emission in 
coincidence with gamma ray bursts (GRBs) to lower energies.
According to \cite{GRB} GRBs may emit a burst
of neutrinos. However, predicted energies are only a few GeV
%below the anticipated physics threshold 
and the  event numbers are small ($\sim 10$\,yr$^{-1}$\,km$^{-2}$).
Additional studies are required to evaluate the sensitivity for such signals.

\section{Summary and Outlook}

This paper summarizes the enhancement of the physics profile of IceCube 
by the DeepCore detector. The geometry of DeepCore has been optimized
 and construction has started. 
Detailed MC studies and experimental analyses are currently under way 
to optimize and finalize the analysis procedures.
First data from the full detector will be available in spring 2010, the veto will be fully completed latest 2011.

\section*{Acknowledgement}
 This work is supported by the German Ministry for Education and Research (BMBF). For a full acknowledgement see \cite{IC3}.
% Maybe we can skip the acknowledgement ???

\end{document}